\documentclass[showpacs,preprintnumbers,amsmath,amssymb,twocolumn,aps,pre]{revtex4-1}
\usepackage{color}
\usepackage{amsmath}
\usepackage{mathrsfs}
\usepackage{graphicx}
\usepackage{subfig}
\usepackage{epsfig}

\begin{document}

\title{Frequency comb generation through the locking of domain walls in doubly resonant dispersive optical parametric oscillators}
\author{P. Parra-Rivas$^{1,2}$, L. Gelens$^{2}$, T. Hansson$^3$, S. Wabnitz$^{4,5}$, and F. Leo$^1$}
\affiliation{ 
 $^1$OPERA-photonics, Université libre de Bruxelles, 50 Avenue F. D. Roosevelt, CP 194/5, B-1050 Bruxelles, Belgium\\
 $^2$Laboratory of Dynamics in Biological Systems, KU Leuven Department of Cellular and Molecular Medicine, University of Leuven, B-3000 Leuven, Belgium\\
 $^3$Department of Physics, Chemistry and Biology, Linköping University, SE-581 83 Linköping, Sweden\\
 $^4$Dipartimento di Ingegneria dell'Informazione, Elettronica e Telecomunicazioni, Sapienza University of Rome, Via Eudossiana 18, 00184 Rome, Italy\\
 $^5$CNR-INO, Istituto Nazionale di Ottica, Via Campi Flegrei 34, I-80078 Pozzuoli (NA), Italy}


\begin{abstract}
In this letter we theoretically investigate the formation of localized temporal dissipative structures, and their corresponding frequency combs in doubly resonant dispersive optical parametric oscillators. We derive a nonlocal mean field model, and show that domain wall locking allows for the formation of stable coherent optical frequency combs.
\end{abstract}
\maketitle
The formation of optical frequency combs (OFCs) in high-Q driven microresonators with Kerr type nonlinearity has attracted considerable attention in the past ten years \cite{DelHaye_2007,Pasquazi_2018}.
Recent works show that dispersive cavities with quadratic nonlinearities
may provide an alternative to Kerr cavities for the generation of frequency combs \cite{Ricciardi_2015,Leo_2016a,Leo_2016b}.
The main advantages of quadratic OFCs are the reduced pump power requirements and the possibility of comb generation in spectral regions that are separate from that of the pump laser frequency. For example, optical parameter oscillators (OPOs) may allow for the efficient generation of OFCs in the mid-infrared using near-infrared continuous wave (CW) sources.
It was recently shown that modulation instability (MI) induces pattern and frequency comb formation in degenerate OPOs \cite{Mosca_PRL_2018}. While promising, it is still unclear whether solitary waves exist in that configuration. 

In this letter, we propose the locking of domain walls (DWs), also called wave fronts and switching-waves, as an alternative mechanism to MI for the generation of OFCs in temporal degenerate OPOs.
A DW consists of a transition connecting two different but coexisting CWs. DWs are particle-like states that can exist separately, interact, and lock forming localized structures (LSs), i.e. domains of finite size that are bi-asymptotic to the CW state \cite{Coullet_1987}.
In the context of Kerr combs, the mechanism of DW locking has been widely studied both experimentally \cite{kippenberg_dark,Xiaoxiao,Garbin_2017}, and theoretically  \cite{PPR_dark1,PPR_dark2}. More recently, it was shown that DWs may form between two polarisation states in fiber resonators \cite{Coen_domain_walls}.  DWs stem from the pi phase indeterminacy of the field above threshold. Both solutions may coexist, forming a wall between states of different phases \cite{Trillo_1997}.
DWs, their interaction and the formation of LSs has been studied, in the absence of walk-off, in the context of {\it spatial diffractive} OPOs, where they arise in the transverse plane to the propagation direction \cite{Oppo_semiclassic_1999, Oppo_2001}. In this work, we show that the same mechanisms can generate similar type of LSs in {\it temporal dispersive} OPOs even when a large temporal walk-off is present. In this case DWs and LSs arise along the propagation direction. To do so we first derive a mean-field model with a nonlocal nonlinear interaction term, and we demonstrate that the formation of the LSs can be easily understood within the framework of a single Ginzburg-Landau equation. We use the system parameters of a recent experiment, demonstrating MI-induced OFCs in singly resonant OPOs \cite{Mosca_PRL_2018}. 

To start, let us consider a dispersive cavity with a quadratic medium phase-matched for degenerate OPO, and driven by the field $B_{in}$ at frequency $2\omega_0$ in a doubly resonant configuration. Such a system can be described by an infinite map for the 
slowly varying envelopes of the fields $A_m(z,t)$ and $B_m(z,t)$, of the electric field 
\begin{equation}
E_m(z,t)={\rm Re}\left[A_m(z,t)e^{i(k_1z-\omega_0t)}+B_m(z,t)e^{i(k_2z-2\omega_0t)}\right],
\end{equation}
centered at frequencies $\omega_0$ and $2\omega_0$, respectively. Propagation of these cavity fields over the $m^{\rm th}$ round trip is governed by
the evolution equations: 
\begin{figure}[t]
	\centering
	\includegraphics[scale=0.72]{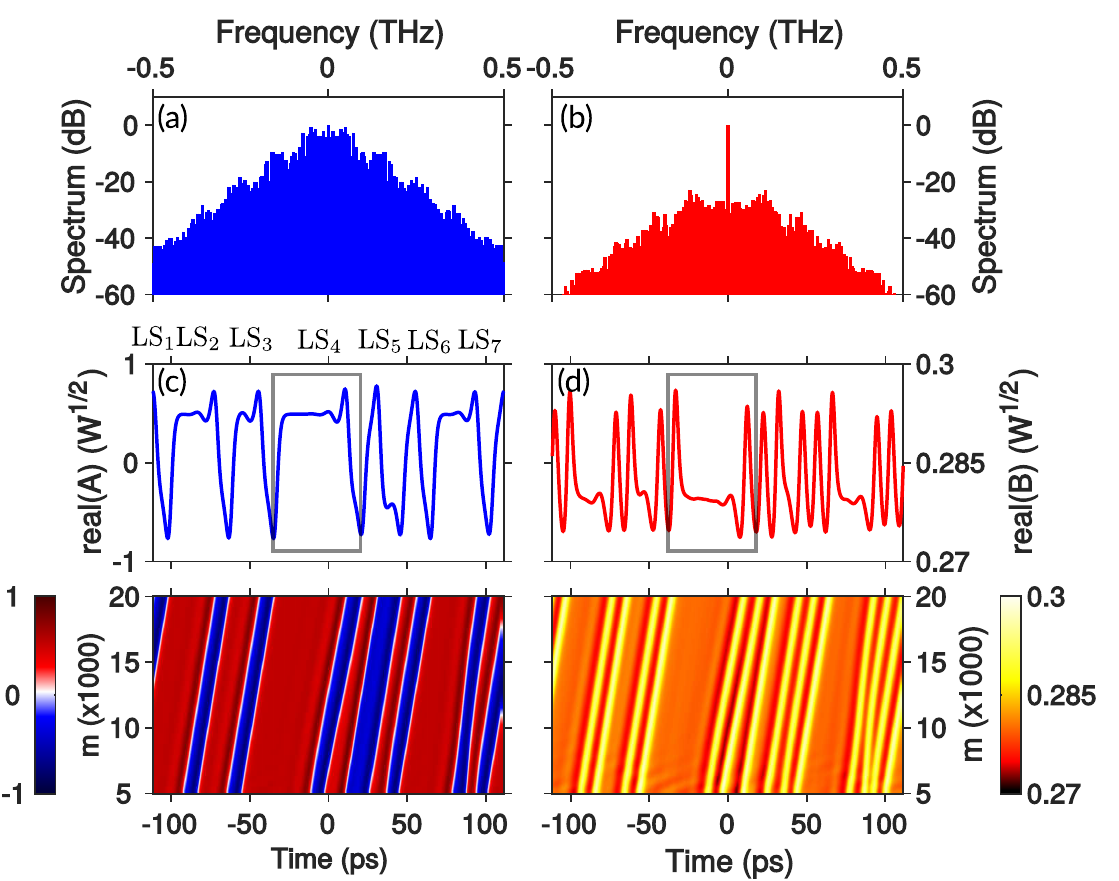}
	
	
	\caption{(a)-(b) show the stationary quadratic OFC obtained from initial random noise after $m=2\cdot 10^4$ cavity round-trips. (c)-(d)[top] show Re$[A_m(0,\tau)]$ and Re$[B_m(0,\tau)]$ of the underlying dissipative structure corresponding to the combs of (a)-(b). In (c)-(d)[bottom] the evolution of these fields is shown after each round-trip. Physical parameters used:  $\kappa=6.58$ W$^{-1/2}$/m, $\alpha_{1,2}=T_{1,2}=0.0196$, $k_1''\approx 0.234$ ps$^2$/m, $k_2''\approx 0.714$ ps$^2$/m, $\Delta k'\approx792$ ps/m, $\delta_1=-0.0392$, $B_{\rm in}=0.395$ W$^{1/2}$, $L_{\rm c}=0.015$ m, and $t_R\approx 223$ ps \cite{Mosca_PRL_2018}.}
	\label{f1}
\end{figure}
\begin{subequations}\label{map1}
	\begin{equation}
	\partial_z A_m=-\left(\displaystyle\frac{\alpha_{c1}}{2}+i\displaystyle\frac{k_1''}{2}\partial_\tau^2\right)A_m+i\kappa B_m\bar{A}_me^{-i\Delta k z}
	\end{equation}
	\begin{equation}
	\partial_z B_m=-\left(\displaystyle\frac{\alpha_{c2}}{2}+\Delta k'\partial_{\tau}+i\displaystyle\frac{k_2''}{2}\partial_\tau^2\right)B_m+i\kappa A_m^2e^{i\Delta k z},
	\end{equation}
\end{subequations}
where $\tau=t-z/v_g$, and $v_g=1/k'_1$ is the group velocity of the fundamental field $A_m$. 
Here $z\in[0,L_c]$ with $L_c$ the length of the cavity; $\alpha_{c1,2}$ describe linear propagation losses for the fields $A_m$ and $B_m$; $k_{1,2}''=d^2k/d\omega^2|_{\omega_0,2\omega_0}$ are the group velocity dispersion (GVD) coefficients; $\Delta k=2k(\omega_0)-k(2\omega_0)$ is the wave-vector mismatch at the degeneracy point; $\Delta k'=dk/d\omega|_{2\omega_0}-dk/d\omega|_{\omega_0}$
is the corresponding group-velocity mismatch, or rate of temporal walk-off; and the nonlinear coupling strength $\kappa\propto\chi^{(2)}$ is normalized such that $|A_m|^2$, $|B_m|^2$ and $|B_{in}|^2$ are measured in Watts. 
Furthermore, the intracavity fields $A_{m+1}(0,\tau)$ and  $B_{m+1}(0,\tau)$ at the beginning of the $(m+1)^{\rm th}$ round-trip are related to the fields at the end of the $m^{\rm th}$ round-trip by
\begin{subequations}\label{map2}
	\begin{equation}\label{BC1}
	A_{m+1}(0,\tau)=\sqrt{1-T_1}A_m(L_c,\tau)e^{-i\delta_1}
	\end{equation}
	\begin{equation}\label{BC2}
	B_{m+1}(0,\tau)=\sqrt{1-T_2}B_m(L_c,\tau)e^{-i\delta_2}+\sqrt{T_2}B_{in},
	\end{equation}
\end{subequations}
where $T_1$ and $T_2$ are the power transmission coefficients, at $\omega_0$ and $2\omega_0$, of the coupler used to inject the CW field $B_{in}$, and $\delta_j=(\omega_j-\omega_{c,j})t_R$ with $j=1,2$ represents the phase detuning of the intracavity field
$A_m$ ($B_m$) for the cavity resonance frequency $\omega_{c_1}$ ($\omega_{c_2}$) closest to $\omega_1=\omega_0$ ($\omega_2=2\omega_0$), over one round-trip time $t_R$ of $A_m$. In what follows we set $\Delta k=0$, $\delta_2=2\delta_1$ \cite{Leo_2016b}.
We use the map (\ref{map1}-\ref{map2}) to numerically  explore the natural dynamics of the system. 
Figs.~\ref{f1}(a) and (b) show the final steady OFCs for $A_m$ and $B_m$  obtained from the evolution of an initial noisy background after a sufficient number of round-trips $(m=2\cdot 10^4)$. These combs correspond to the temporal dissipative structures shown in panels (b) and (c)[top], where
Re[$A_m$] and Re[$B_m$] are plotted in blue and red, respectively.	
The evolution of the fields after every round-trip is shown in the bottom sub-panels of (c)-(d). 
Both fields exhibit a constant temporal drift of about  $0.0014$ ps per round-trip. Looking to the $A_m$ field [see Fig.~\ref{f1}(c), top], we can identify a sequence of DWs connecting two different continuous wave states, forming a disordered stationary state.
This particular solution consists of seven LSs (see LS$_1$ - LS$_7$) of different widths and separations between them.  At the location of each of the DWs in the $A_m$ field, one finds pulses in the $B_m$ field [see Figure~\ref{f1}(d), top]. Furthermore, one can use each such LS [see for instance LS$_4$] as a new initial condition for the map, and find that each LS is also a localized steady state solution of the system. Moreover, when starting with a noisy background and the same parameter set, we find that different realizations of the noise can lead to many different sequences of such pulses, revealing multistability of LSs.
%
%
%
As far as we know, the existence of this type of structures has not yet been reported in the context of quadratic dispersive cavities. Therefore, it is important to elucidate their formation and properties. For doing so, we first derive a single time-domain equation that allows for the description of stationary states associated with LSs, and their spectral properties.

Assuming that the resonator exhibits high finesse, that both fields do not vary significantly over a single round-trip (i.e., the combined effects of nonlinearity and dispersion are weak), and following Refs.~\cite{Haelterman,Leo_2016b,Hansson_2017}
one can reduce Eqs.~(\ref{map1}) and (\ref{map2}), to two coupled mean-field equations:
\begin{subequations}\label{MF}
	\begin{equation}\label{MF1}
	\partial_t A=-(1+i\Delta_1)A-i\beta_1\partial_{\tau'}^2A+i B\bar{A}
	\end{equation}
	\begin{equation}\label{MF2}
	\partial_t B=-(\alpha+i\Delta_2)B-\left(d\partial_{\tau'}+i\beta_2\partial_{\tau'}^2\right)B+ iA^2 +S,
	\end{equation}
\end{subequations}
where $A(t,\tau)$ and $B(t,\tau)$ are the normalized cavity field envelopes at $z=0$ defined as 
$A(t=mt_R,\tau)=\kappa L_cA_m(z=0,\tau)/\alpha_1$ and 
$B(t=mt_R,\tau)=\kappa L_cB_m(z=0,\tau)/\alpha_1$,
and the index $m$ has been replaced by the slow-time variable $t=mt_R$.
Here the normalized variables and parameters are
$\alpha=\alpha_2/\alpha_1$, where $\alpha_{1,2}=(T_{1,2}+\alpha_{c1,2}L_c)/2$, $\Delta_{1,2}=\delta_{1,2}/\alpha_1$, $d=\Delta k'\sqrt{2L_c/\alpha_1|k_1''|}$, $S=B_{in}\sqrt{T_2}\kappa L_c/\alpha_1^2$, $\beta_1={\rm sign}(k_1'')$, $\beta_2=k_2''/|k_1''|$, and $\tau'=\tau\sqrt{2\alpha_1/|k_1''|L_c}$.

Interestingly, we find through numerical inspection that, for a large range of parameters, $B$ evolves slowly in $t$. Thus, we can make the approximation that the term $\partial_tB$ can be neglected in Eq.~(\ref{MF2}). Under this observation we can further simplify Eqs.~(\ref{MF}) to a single mean-field model, as done in Refs.~\cite{Nikolov_2003,Hansson_2017}. 
To do so, from Eq.~(\ref{MF2}) one can obtain an expression of $B$ as a function of $A^2$ and $S$, that once inserted in 
Eq.~(\ref{MF1}) gives: 
\begin{equation}\label{nl_GL}
\partial_t {\mathsf A}=-(1+i\Delta_1)\mathsf{A}-i\beta_1\partial_{\tau'}^2\mathsf{A}-\bar{\mathsf{A}}(\mathsf{A}^2\otimes\mathsf{J}) +\rho\bar{\mathsf{A}}.
\end{equation}
with $\otimes$ denoting the convolution with the nonlocal kernel 
\begin{equation}
\mathsf{J}(\tau')=\frac{1+\tilde{\Delta}_2^2}{2\pi}\int_{-\infty}^{\infty}\frac{e^{-i\Omega\tau'}d\Omega}{1+i(\tilde{\Delta}_2-\gamma \Omega-\eta \Omega^2)},
\end{equation}
where $\tilde{\Delta}_2=\Delta_2/\alpha$, $\gamma=d/\alpha$, $\eta=\beta_2/\alpha$. The term $\mathsf{A}^2\otimes\mathsf{J}$ introduces a nonlocal nonlinear coupling between the points of the fast variable $\tau$ \cite{Clerc_2005,Lendert_2010}. The normalized field reads $\mathsf{A}=Ae^{-i\psi}/\sqrt{\alpha(1+\tilde{\Delta}_2^2)}$, with $\psi=\pi/4+{\rm atan}(-\tilde{\Delta}_2)/2$, and the normalized driving field amplitude is $\rho= S/(\alpha\sqrt{1+\tilde{\Delta}_2^2})$. 
With this approximation, the $B$ field is dynamically slaved to the $A$ field, and it is explicitly given by
$B=-(\mathsf{A}\otimes\mathsf{J}+\rho)e^{i{\rm atan}(\tilde{\Delta}_2)}$.  Without loss of generality we will consider the normal GVD regime ($\beta_1=1$), and $\alpha=1$, such that $\tilde{\Delta}_2=2\Delta_1$.

We have used a variety of steady state solutions obtained with the infinite map (\ref{map1})-(\ref{map2}), such as those in Fig. \ref{f1}, as initial condition in the two models (\ref{MF}) and (\ref{nl_GL}). We found that these initial temporal profiles quickly converge, and that the converged solutions were almost identical in all models. 
This confirms the validity of both mean field models (\ref{MF}) and (\ref{nl_GL}).

\begin{figure}[!t]
	\centering
	\includegraphics[scale=1.1]{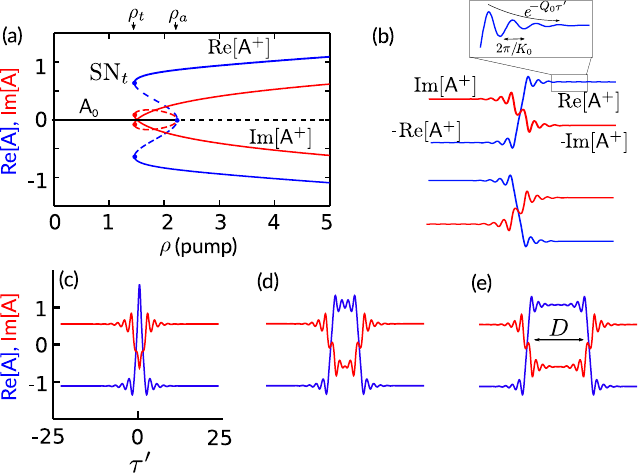}
	\caption{In (a) we show the HSS solution of Eq.~(\ref{nl_GL}) in the subcritical regime for $\Delta_1=-2$. Panel (b) shows an example of two DW solutions of Eq.~(\ref{GL}) with different polarities for  $(\Delta_1,\rho)=(-2,4)$. For the same parameters, panels (c)-(e) show different LSs formed through the locking of the DWs at different separations. Here $(\gamma,\eta)=(0,0)$.}
	\label{f2}
\end{figure}
Based on our observations in Fig.~\ref{f1}(c)[top], we anticipate that the LSs are composed of two DWs connecting different CWs that coexist for the same parameter values and that are related by the transformation $\mathsf{A}\rightarrow -\mathsf{A}$. The CWs correspond to the homogeneous steady state (HSS) solutions of Eq.~(\ref{nl_GL}), namely 
$\mathsf{A}_0=0$, and the two branches of solutions $\mathsf{A}=\mathsf{A}^{\pm}$
satisfying
\begin{equation}\label{hom}
|\mathsf{A}^{\pm}|^2=\frac{(\Delta_1\tilde{\Delta}_2-1)\pm\sqrt{(1+\tilde{\Delta}_2^2)\rho^2-(\tilde{\Delta}_2+\Delta_1)^2}}{1+\tilde{\Delta}_2^2},
\end{equation}
with 
$\mathsf{A}^{\pm}=|\mathsf{A}^{\pm}|e^{i\phi^\pm}$, and 
$\phi^\pm={\rm acos}\left[(|\mathsf{A}^{\pm}|^2+1)/\rho\right]/2$.
Here we only consider DWs between $-\mathsf{A}^+$ and $\mathsf{A}^+$, as done in Refs.~\cite{Oppo_semiclassic_1999,Oppo_2001}.
For large enough detuning, $\Delta_1>1/\tilde{\Delta}_2$, only the $\mathsf{A}^+$ branch exists, and bifurcates supercritically from a pitchfork bifurcation at a pump strength $\rho_a=\sqrt{1+\Delta_1^2}$. However, if 
$\Delta_1<1 / \tilde{\Delta}_2$, $\mathsf{A}^-$ emerges subcritically (and therefore unstably), and stabilizes at a saddle-node SN$_t$, at $\rho_t=(\tilde{\Delta}_2+\Delta_1)/\sqrt{1+\tilde{\Delta}_2^2}$, where it merges with $\mathsf{A}^+$. We will focus on the subcritical regime, whose bifurcation diagram is shown in Fig.~\ref{f2}(a).

When coupling between points of the fast time variable (i.e. $\tau'$) is introduced in the system through the dispersion term $\beta_1\partial_{\tau'}^2\mathsf{A}$ and/or the nonlocal nonlinear coupling $\mathsf{A}^2\otimes \mathsf{J}$, connections between the HSSs can arise.
Furthermore, LSs can form through the locking of these DWs. To illustrate this mechanism, let us first consider the presence of  $\beta_1\partial_{\tau'}^2\mathsf{A}$ only. A natural way for doing so is to neglect walk-off and the GVD of $B$, so that $\gamma=0$, and $\eta=0$. In this case, the convolution $\mathsf{A}^2\otimes \mathsf{J}$ reduces to $(1-i\tilde{\Delta}_2)\mathsf{A}^2$ and Eq.~(\ref{nl_GL}) becomes
\begin{equation}\label{GL}
\partial_t {\mathsf A}=-(1+i\Delta_1)\mathsf{A}-i\beta_1\partial_{\tau'}^2\mathsf{A}-(1-i\tilde{\Delta}_2)|\mathsf{A}|^2\mathsf{A} +\rho\bar{\mathsf{A}},
\end{equation}
which is a simpler version of the well known parametrically forced Ginzburg-Landau equation with 2:1 resonance \cite{Yochelis_2008}. A similar equation can be also derived in the framework of singly resonant diffractive OPOs \cite{Oppo_2001}. Notice that the HSS solutions of this equation are also given by Eq.~(\ref{hom}).
In diffractive systems that are not bounded periodically, DWs connecting the equivalent states $-\mathsf{A}^+$ and $\mathsf{A}^+$ can exist in both supercritical and subcritical regimes \cite{Trillo_1997}. 
Fig.~\ref{f2}(b) shows two examples of a DW in the subcritical regime for $(\Delta_1,\rho)=(-2,4)$, one showing an upwards connection of $-\mathsf{A}^+$ to $\mathsf{A}^+$ (top),
and the other showing a downwards connection from $\mathsf{A}^+$ to $-\mathsf{A}^+$ (bottom). We refer to these two types of DWs as having an opposite polarity.
Furthermore, they are invariant under the simultaneous transformations $\tau'\mapsto-\tau'$, and $\mathsf{A}\mapsto-\mathsf{A}$,
and they are stationary, i.e. they are {\it Ising fronts} \cite{Coullet_1990}.

In order for DWs of opposite polarity to form stable connections, it is of critical importance to consider the way that the wave fronts approach the HSS asymptotically. In the linear regime, this approach can be described by $\mathsf{A}(\tau')=\mathsf{A}^{+}+ a e^{\lambda \tau'}+ c.c.$, where $|a|\ll1$ and $\lambda=Q+iK$, with $Q$ and $K$ real numbers. These eigenvalues $\lambda$ depend on the control parameters of the system, and can be obtained by studying the system of linearized ordinary differential equations for the perturbations, derived through direct substitution of the ansatz into Eq.~(\ref{nl_GL}) \cite{Oppo_2001,Yochelis_2008}. In the linear regime (i.e. far from the DW core), the overall shape of the DW oscillatory tail that approaches the HSS is determined by the leading eigenvalue $\lambda_0=Q_0+iK_0$, which is the eigenvalue with the real part closest to zero, as all the other directions are damped faster.  An example of an oscillatory tail ($K_0 \neq 0$) can be seen in detail in the close-up view of Fig.~\ref{f2}(b). If $\lambda_0$ is purely real, the front approaches the HSS monotonically.


In a periodic system like ours, single isolated DWs do not exist: to satisfy the boundary conditions they must necessarily always come in pairs.
Two DWs of opposite polarity exhibit a particle-like interaction force whose strength decays exponentially with their temporal separation $D$ [see Fig.~\ref{f2}(e)], as
described by $\partial_t D\sim e^{-Q_0D}{\rm cos}(K_0 D) $ \cite{Coullet_1987}.
The stationary solutions of this equation, $D_n\equiv2\pi n/K_0$ (with $n\in\mathbb{N}$), correspond to the locations at which two DWs can lock to each other, and form a LS of width $D\approx D_n$. Due to the oscillatory nature of this force, the interaction of DWs alternates between attraction and repulsion, as does the stability of the stationary separations. Thus, for a given set of parameters, multiple LSs of different widths can coexist, as illustrated in Fig.~\ref{f2}(c)-(e). These LSs only form when the DWs approach the HSS in an oscillatory way ($K_0 \neq0$). In contrast, when the tails are monotonic ($K_0=0$), two DWs attract each other, and move towards each other until they annihilate one another in a process called coarsening \cite{Allen_Cahn}. 
\begin{figure}[!t]
	\centering
	\includegraphics[scale=1]{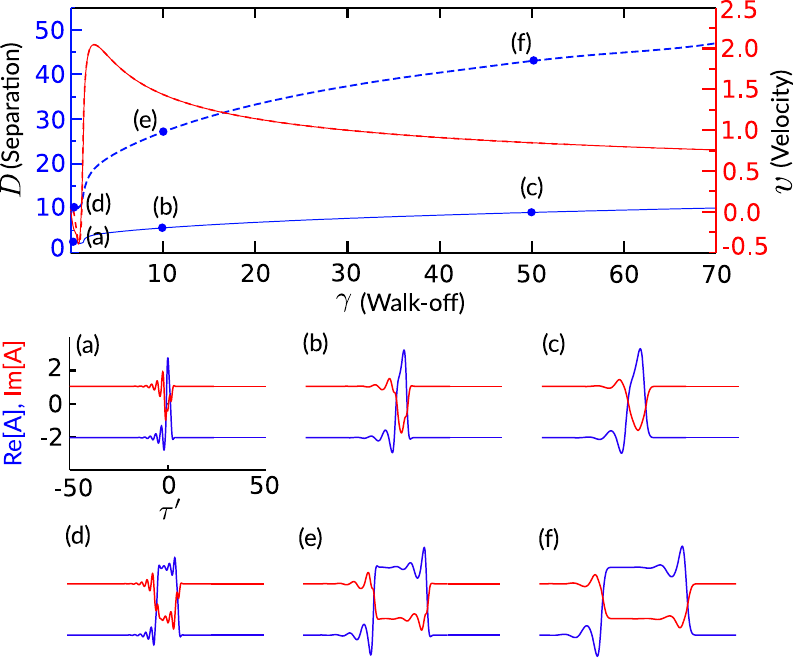}
	\captionsetup{justification=raggedright,singlelinecheck=false}
	\caption{Bifurcation diagram showing the influence of walk-off $\gamma$ on a single bump LS in the absence of $\eta$ (the GVD of $B$). Here the width $D$ and speed $v$ of the LS are shown as a function of $\gamma$. The labels (a)-(d) correspond to the LSs profiles shown in the subpanels. Here $(\Delta_1,\rho)=(-2,4)$.}
	\label{f4}
\end{figure}

When either the GVD of the $B$ field, $\eta$, and/or the walk-off, $\gamma$, are present, the nonlocal nonlinearity must be taken into account, and one has to consider the more general Eq.~(\ref{nl_GL}).
In this model, DW solutions still exist, although their shape and symmetry properties are modified, since nonlocal nonlinear coupling can significantly alter the eigenvalues $\lambda$ \cite{Clerc_2005,Lendert_2010}. Nevertheless, the mechanism of DW locking remains the same for solutions of Eq.~(\ref{nl_GL}), and similar LSs can be found. To illustrate this, we explore the influence of the walk-off $\gamma$ on the LSs that we found using the local model [see Fig.~\ref{f2}].
The walk-off breaks the left/right symmetry ($\tau \mapsto -\tau'$), and the LSs now drift at a constant velocity $v$ proportional to $\gamma$. Considering a change of coordinates to a moving reference frame (i.e. $\tau'\mapsto \tau'-vt$) and using a numerical continuation algorithm, based on a Newton-Raphson solver, the LSs and their velocity $v$ can be continuously tracked in the parameter $\gamma$. In this way, the bifurcation diagram in Fig.~\ref{f4} was constructed, where we tracked two LS of different widths [i.e. the solutions shown in Fig.~\ref{f2}(c) and Fig.~\ref{f2}(d)].
The diagram at the top of Fig.~\ref{f4} shows how the width (blue) and the velocity $v$ (red) of these LS solutions change with increasing walk-off $\gamma$. The solid lines corresponds to the continuation of a narrow LS [Fig.~\ref{f2}(c)], while the dashed lines show the changes of a wider LS [Fig.~\ref{f2}(d)]. Changes in the profiles of the LSs are illustrated in the subpanels (a)-(f). In both cases, the width of the LSs increases monotonically with $\gamma$, due to an increase in the wavelength of the oscillatory tails of the DWs [see the profiles plotted in Fig.~\ref{f4}(a)-(f)].  The velocities of both LSs are approximately equal, indicating that the velocity depends mostly on the strength of the walk-off, and not on the shape and width of the structures. The LS velocity, initially negative, becomes positive at $\gamma\approx2$, and then sharply increases until reaching a peak value $v\approx2$, after which it decreases monotonically and saturates for large values of the walk-off $\gamma$. For large walk-off, the LSs of Fig.~\ref{f4} closely resemble the structures that we found by simulating the full map, as shown in Fig.~\ref{f1}. The fact that 
this type of LSs exists for very large values of walk-off can have a big practical advantage for OFC generation since no dispersion engineering of the resonator
is needed other than for the phase-matching.
By incorporating both walk-off $\gamma$ and GVD $\eta$, we can then retrieve exactly the LSs in Fig.~\ref{f1}.
In the absence of $\gamma$, the presence of $\eta$ induces modifications in the tails of the LSs.
However, for the parameter values considered in this work, $\gamma$ dominates and the effect of $\eta$ is almost negligible. 
The influence of the different control parameters of the system on the bifurcation structure of the LSs can be quite complex and their study is beyond the scope of this paper. 


In summary, we have studied the formation of temporal LSs in doubly resonant dispersive OPOs in the presence of walk-off. Using an infinite map for the slowly varying envelopes of the OPO fields, we showed that a random sequence of LSs of different widths is formed naturally. We then derived a nonlocal mean-field model with a nonlocal nonlinear coupling that describes the dynamics of these OPO cavities, in which we confirmed the existence of LSs.
They are formed through the locking of DWs connecting two equivalent homogeneous states, which could be explained more easily
in the local case. Afterwards, using continuation techniques, we studied how these LSs are altered by walk-off and GVD of the pump field $B$. Remarkably, the formation of this type of LSs does not depend on modulational instabilities.
In the frequency domain, LSs correspond to coherent frequency combs formed around both $\omega_0$ and $2\omega_0$. We expect that these structures may play a significant role for future integrated ultra-broadband frequency comb generation.

\textbf{Acknowledgment.}
We acknowledge the support from internal Funds from KU Leuven and the FNRS (PPR); the Ministero dell’ Istruzione, dell’ Università e della Ricerca (MIUR), [Grant 2015KEZNYM], (SW); funding from the Swedish Research Council, [Grant No. 2017-05309], (TH); funding from the European Research Council (ERC) under the European Union’s Horizon 2020 research and innovation programme [Grant agreement No. 757800], (FL).

\end{document}